\documentclass[a4paper,11pt]{article}
\usepackage{pos}

\usepackage{graphicx}
\usepackage{blindtext}
\usepackage{color}
\usepackage{hyperref}
\usepackage{adjustbox}
\usepackage{lineno}
\title{Measurements of direct-photon production in \PbPb collisions at \fivenn and \twosevensixnn with~the ALICE experiment}
\ShortTitle{Direct-photon production in \PbPb collisions at LHC }

\author*[a]{A. Marin}
\onbehalf{on behalf of the ALICE Collaboration}

\affiliation[a]{GSI Helmholtzzentrum f\"ur Schwerionenforschung GmbH, \\
Planckstrasse 1, 64291 Darmstadt, Germany}

\emailAdd{a.marin@gsi.de}

\abstract{Recent results on direct-photon measurements in \PbPb collisions at \fivenn from central to peripheral collisions,  as well as  in 0--10\% central and 20--40\%  semicentral \PbPb collisions 
at \twosevensixnn with improved significance  are presented. 
A significant direct-photon signal is measured for \pT $\gtrapprox$ 2 \GeVc from central to peripheral  \PbPb collisions at \fivenn which is in agreement 
with model calculations  containing pre-equilibrium photons in the intermediate \pT range and prompt photons at high \pT.  No significant direct-photon signal is measured in the low \pT  interval which is also
in agreement with the small thermal-photon signal predicted by state-of-the art models. 
On the other hand, a direct-photon signal is measured for \pT $>$ 1 \GeVc with a significance of 3.1 (1.0 $<$ \pT $<$ 1.8 \GeVc) and 3.4 (1.0 $<$ \pT $<$ 2.3 \GeVc) in 0--10\%  and 20--40\% central \PbPb collisions at \twosevensixnn, respectively.
The inverse slope parameters of non-prompt photon spectra in the low \pT range, 1.1 \GeVc $<$ \pT $<$ 2.1 \GeVc, are
$T_{\rm eff} = 343 \pm 32 (stat)\pm 68(sys)$ MeV and   $T_{\rm eff}  = 339 \pm 32 (stat)\pm 68(sys)$ MeV for central and semicentral collisions. In  the  intermediate \pT range, 2.1 \GeVc $<$ \pT $<$ 4.0 \GeVc, the effective temperatures
are $T_{\rm eff}  = 406 \pm 19 (stat)\pm 36(sys)$ MeV and   $T_{\rm eff}  = 458 \pm 25 (stat)\pm 40(sys)$ MeV, respectively. 
The effective temperatures  $T_{\rm eff}$ in the intermediate \pT range are systematically larger compared to those obtained at low \pT (although consistent within uncertainties) which may indicate an 
earlier photon emission, and thus sensitivity to pre-equilibrium photon production.
Direct photons at both energies are in agreement with state-of-the-art theory calculations over the complete \pT range and measured centralities.}

\FullConference{ HardProbes2023\\
 26-31 March 2023 \\
 Aschaffenburg, Germany\\}

\newcommand{\PbPb}         {\mbox{Pb--Pb}\xspace}

\newcommand{\AuAu}         {\mbox{Au--Au}\xspace}

\newcommand{\sNN}          {\ensuremath{\sqrt{s_{\mathrm{NN}}}}\xspace}
\newcommand{\pT}           {\ensuremath{p_{\rm T}}\xspace}

\newcommand{\nineH}        {$\sqrt{s}~=~0.9$~Te\kern-.1emV\xspace}
\newcommand{\seven}        {$\sqrt{s}~=~7$~Te\kern-.1emV\xspace}
\newcommand{\twoH}         {$\sqrt{s}~=~0.2$~Te\kern-.1emV\xspace}
\newcommand{\twosevensix}  {$\sqrt{s}~=~2.76$~Te\kern-.1emV\xspace}
\newcommand{\five}         {$\sqrt{s}~=~5.02$~Te\kern-.1emV\xspace}
\newcommand{\twoHnn}         {$\sqrt{s_{\mathrm{NN}}}~=~0.2$~Te\kern-.1emV\xspace}
\newcommand{\twosevensixnn}{$\sqrt{s_{\mathrm{NN}}}~=~2.76$~Te\kern-.1emV\xspace}
\newcommand{\fivenn}       {$\sqrt{s_{\mathrm{NN}}}~=~5.02$~Te\kern-.1emV\xspace}
\newcommand{\fivennFour}       {$\sqrt{s_{\mathrm{NN}}}~=~5.44$~Te\kern-.1emV\xspace}
\newcommand{\fivennThree}       {$\sqrt{s_{\mathrm{NN}}}~=~5.36$~Te\kern-.1emV\xspace}

\newcommand{\GeVc}         {Ge\kern-.1emV/$c$\xspace}
\newcommand{\MeVc}         {Me\kern-.1emV/$c$\xspace}
\newcommand{\TeV}          {Te\kern-.1emV\xspace}
\newcommand{\GeV}          {Ge\kern-.1emV\xspace}
\newcommand{\MeV}          {Me\kern-.1emV\xspace}
\newcommand{\GeVmass}      {Ge\kern-.2emV/$c^2$\xspace}
\newcommand{\MeVmass}      {Me\kern-.2emV/$c^2$\xspace}

\newcommand{\VZERO}        {\rm{V0}\xspace}

\newcommand{\piz}{\ensuremath{\pi^{0}}\xspace}
\newcommand{\e}{\ensuremath{\eta}\xspace}

\newcommand{\Rg}{\ensuremath{R_{\rm{\gamma}}}\xspace}

\newcommand{\mT}           {\ensuremath{m_{\rm T}}\xspace}

\begin{document}
\maketitle

\section{Introduction}
Direct photons are unique probes to study and characterize the quark-gluon plasma (QGP) as they leave the medium not affected by strong interactions. They are produced throughout  all stages of the collision; thus,
they carry information about the space-time evolution and  the temperature of the medium. 
Direct photons are all photons except the ones from hadron decays. They include several sources \cite{David:2019wpt}. The prompt photons produced in initial hard scatterings are dominant at high \pT (\pT $>$ 4 \GeVc).
Their production can be described by next-to-leading order (NLO) perturvative Quantum Chromodynamics (pQCD) calculations. Next in production time are pre-equilibrium photons, that can be found in the 
intermediate \pT range (2 $<$\pT $<$ 4 \GeVc).
Thermal photons from the QGP and hadron gas phase are the dominant contribution at low \pT \cite{vanHees:2014ida,Paquet:2015lta,Dasgupta:2018pjm,Linnyk:2015tha,Gale:2021emg}. 
They can be calculated employing state-of-the-art hydrodynamical calculations. 
There are other contributions like photons produced in jet-medium interactions. The spectrum  evolves from an exponential shape  at low \pT to a power-law shape at high \pT.

Direct-photon production was measured in \PbPb collisions at \twosevensixnn in three
centrality classes by combining PCM and PHOS measurements in ALICE \cite{Adam:2015lda}. The significance of the measurement at low \pT was largely influenced by the 4.5\% material budget uncertainty of the PCM measurement. 
A  recent data-driven precision determination of the material budget in ALICE  \cite{ALICE:2023kzv}  which reduces the uncertainty down to 2.5\% is 
used for the direct-photon measurements in \PbPb collisions at \twosevensixnn and \fivenn  presented in these proceedings from the 2011 and 2015 data samples, respectively.   


\section{Analysis methods}
Photons can be detected in ALICE \cite{Abelev:2014ffa} using the PHOS \cite{ALICE:2019cox} and the EMCal \cite{ALICE:2022qhn} calorimeters, or via the photon conversion method (PCM) \cite{Abelev:2014ffa}. 
In PCM, photons  are measured by reconstructing the electron positron pairs
produced by photon conversions using the ITS \cite{Aamodt:2010aa}  and the TPC \cite{Alme:2010ke} detectors located in the central barrel of the ALICE experiment inside the L3 magnet.
The \VZERO detectors  and Zero Degree Calorimeters (ZDC) are used for triggering and characterization of the collision. In the 2011 run, an online  centrality trigger based on the amplitude of the \VZERO detectors was used
to enhance the central and semicentral events, while during the 2015 run a minimum bias (MB) condition was used for the complete data sample. The data samples consist of a total of 75 million MB \PbPb collisions for the 2015 run, 
and  20 million 0--10\% central and 8.4 million  20--40\% semicentral collisions for the 2011 run. The PCM only measurement is used for the 2015 direct-photon analysis, 
and the PCM  analysis from 2011  is combined with the PHOS 2010 analysis  to obtain the improved results in \PbPb collisions at \twosevensixnn.

\section{Direct photons in \PbPb collisions at \twosevensixnn and at \fivenn}
\begin{figure}[htb]
\begin{center}
\begin{tabular}{lr}
  \includegraphics[width=0.48\linewidth,trim={0.5cm 4.cm 0.5cm 4.2cm }, clip]{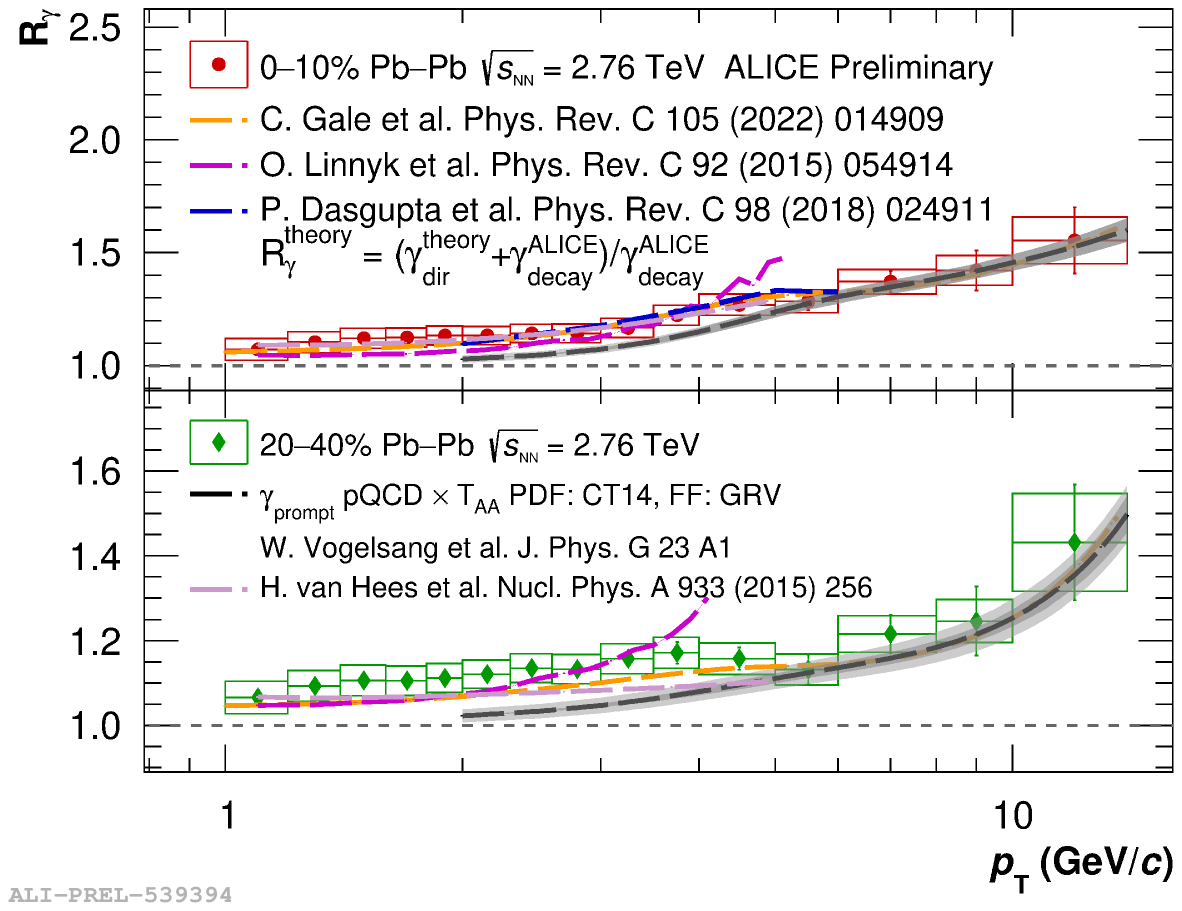}&
  \includegraphics[width=0.5\linewidth,trim={0.5cm 4.cm 0.5cm 4.2cm }, clip]{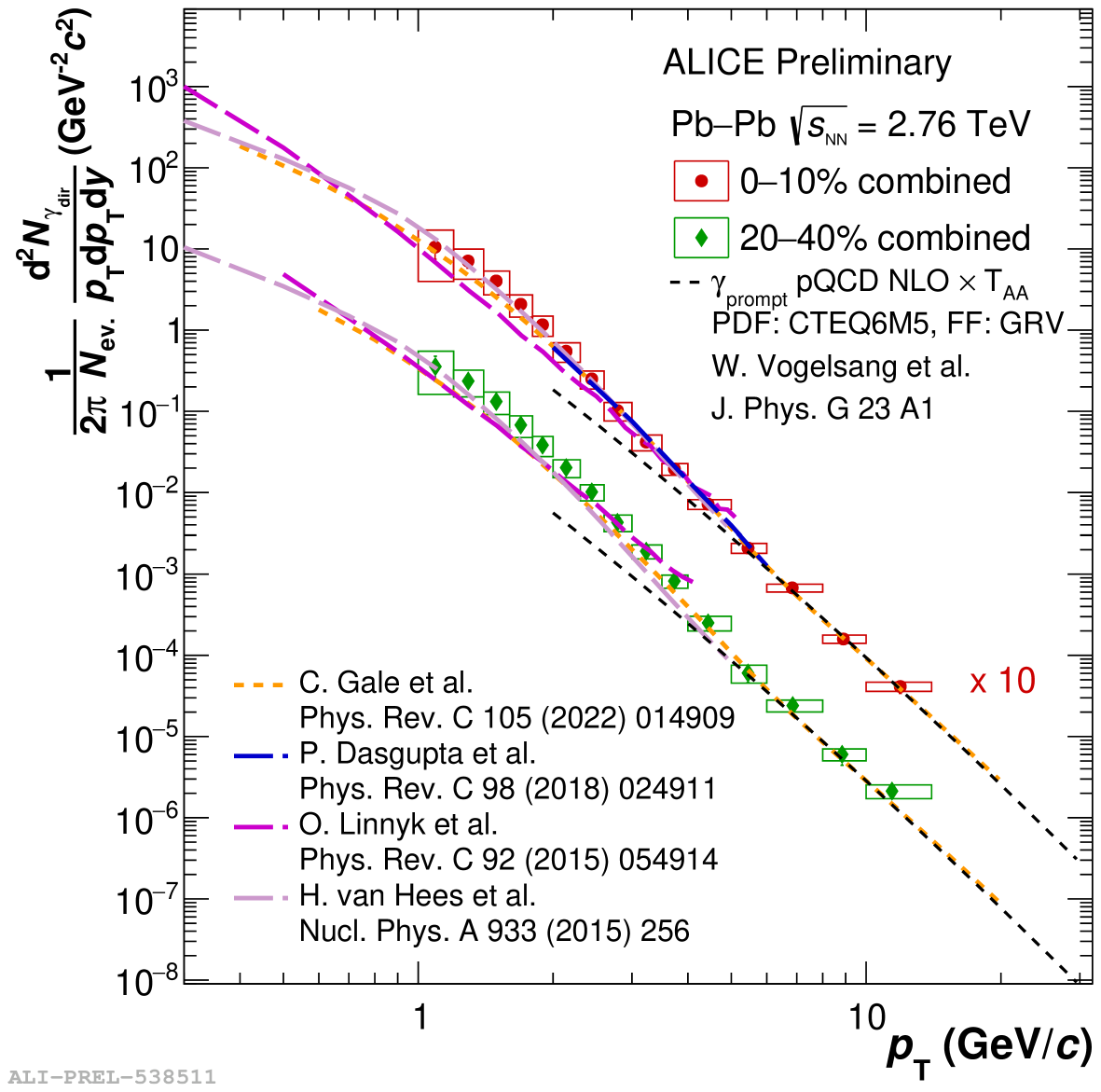}\\
  \end{tabular}
\end{center}
 \caption{\Rg (left)  and direct-photon spectra (right) as a function of \pT for 0--10\% central (top)  and   20--40\%  semicentral (bottom) \PbPb collisions at \twosevensixnn compared different model calculations \cite{vanHees:2014ida,Paquet:2015lta,Dasgupta:2018pjm,Linnyk:2015tha,Gale:2021emg}. }  
 \label{fig:Rgamma276}
\end{figure}
\begin{figure}[htb]
\begin{center}
\begin{tabular}{cc}
\includegraphics[width=0.46\linewidth,trim={0.cm 0.cm 0.cm 0.2cm}, clip ]{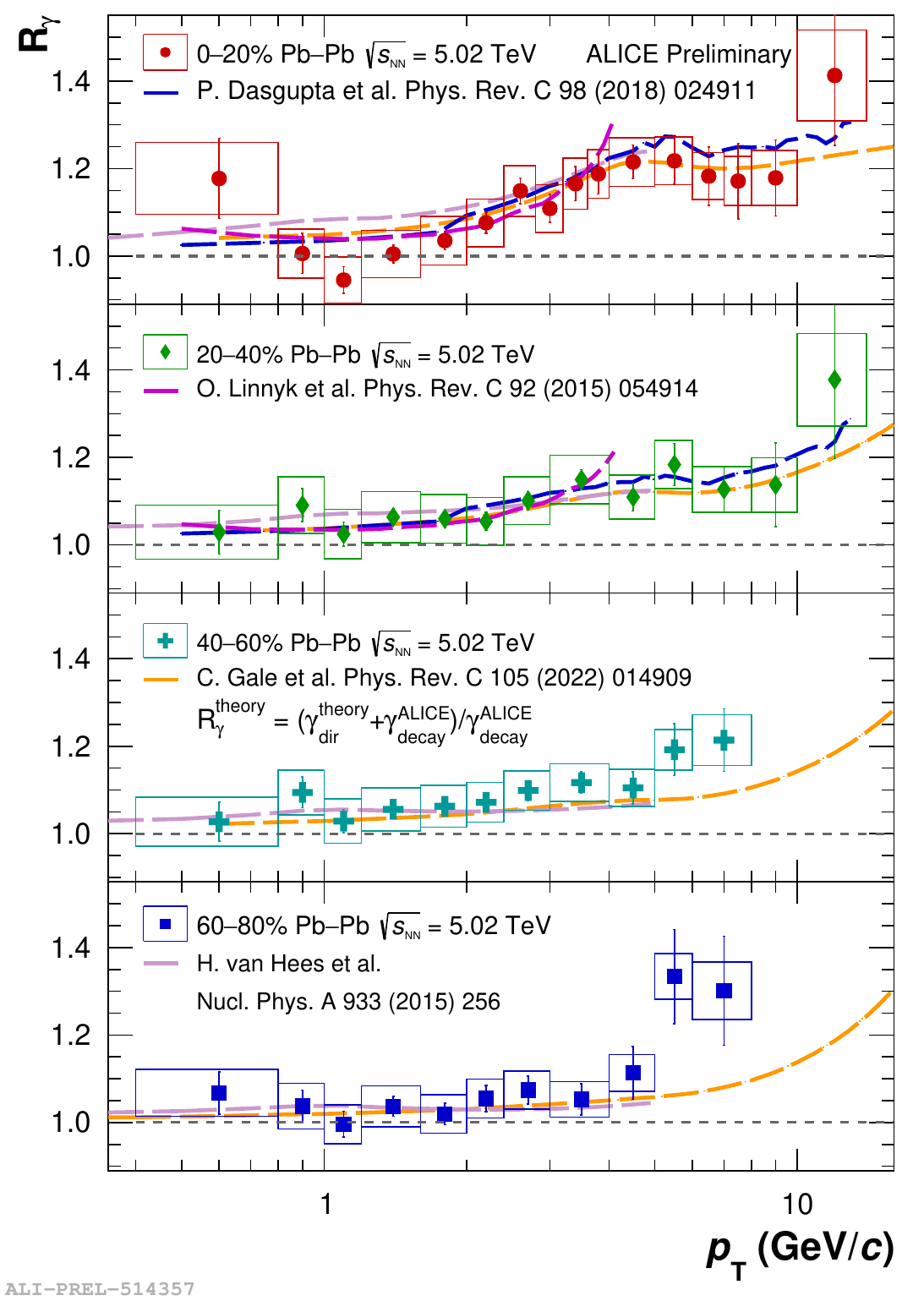}&
\includegraphics[width=0.46\linewidth,trim={0.cm 0.cm 0.cm 0.2cm}, clip ]{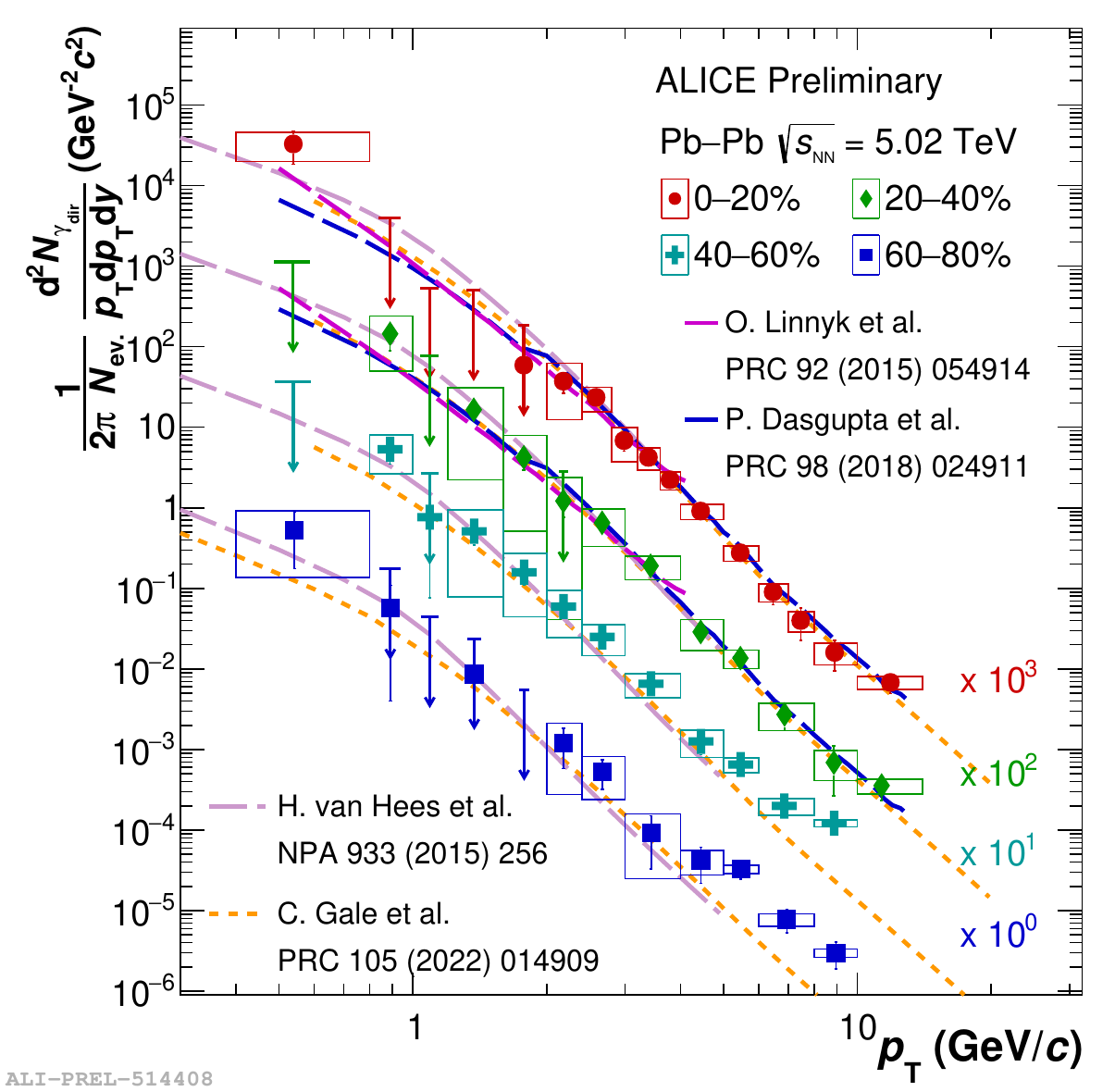}\\
   \end{tabular}
\end{center} 
 \caption{\Rg (left) and direct-photon spectra (right) as a function of \pT from central (top) to peripheral (bottom) \PbPb collisions at \fivenn compared to different  model calculations \cite{vanHees:2014ida,Paquet:2015lta,Dasgupta:2018pjm,Linnyk:2015tha,Gale:2021emg}. }
  \label{fig:Rgamma502}
\end{figure}
\begin{figure}[htb]
\begin{center}
\hspace*{-0.35cm}\begin{tabular}{cc}
\includegraphics[width=0.46\linewidth,trim={0.5cm 6.7cm 0.5cm 6.7cm}, clip] {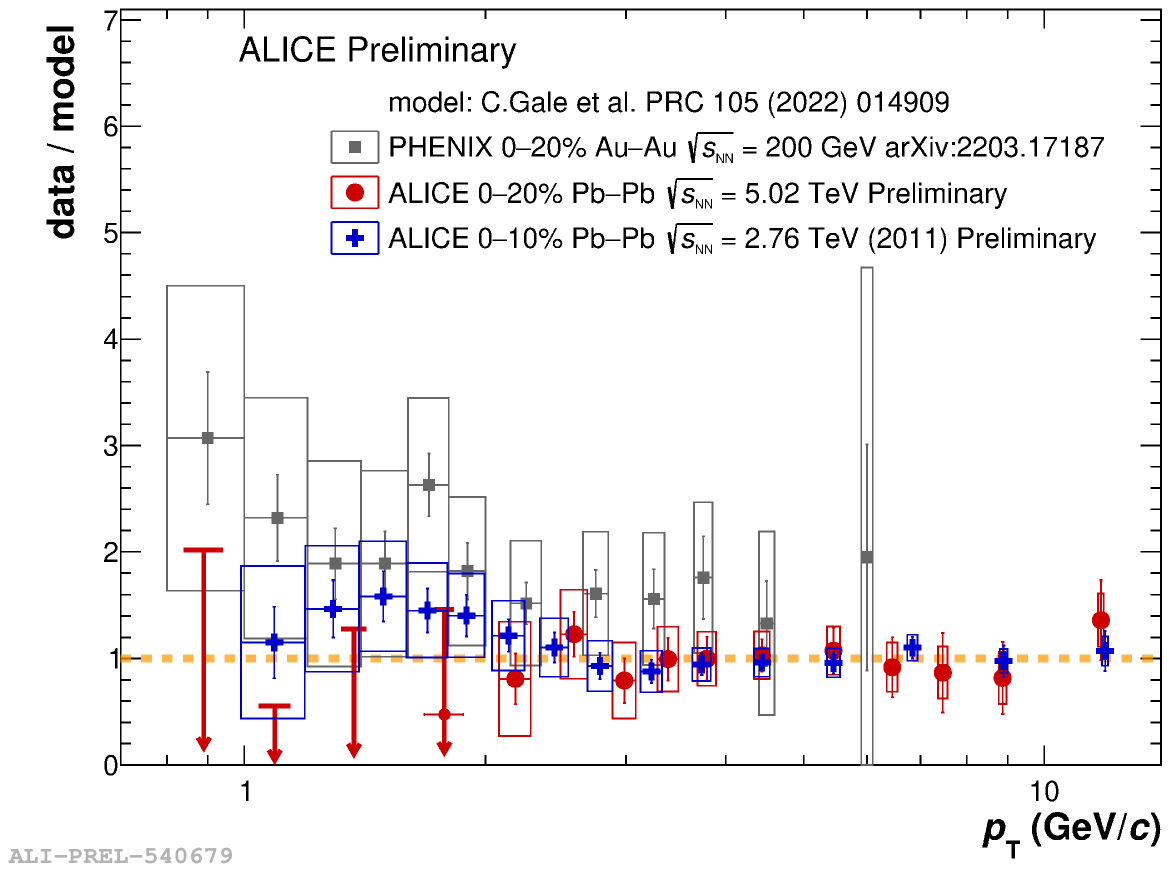}&
\includegraphics[width=0.46\linewidth,trim={0.5cm 6.7cm 0.5cm 6.7cm}, clip ]{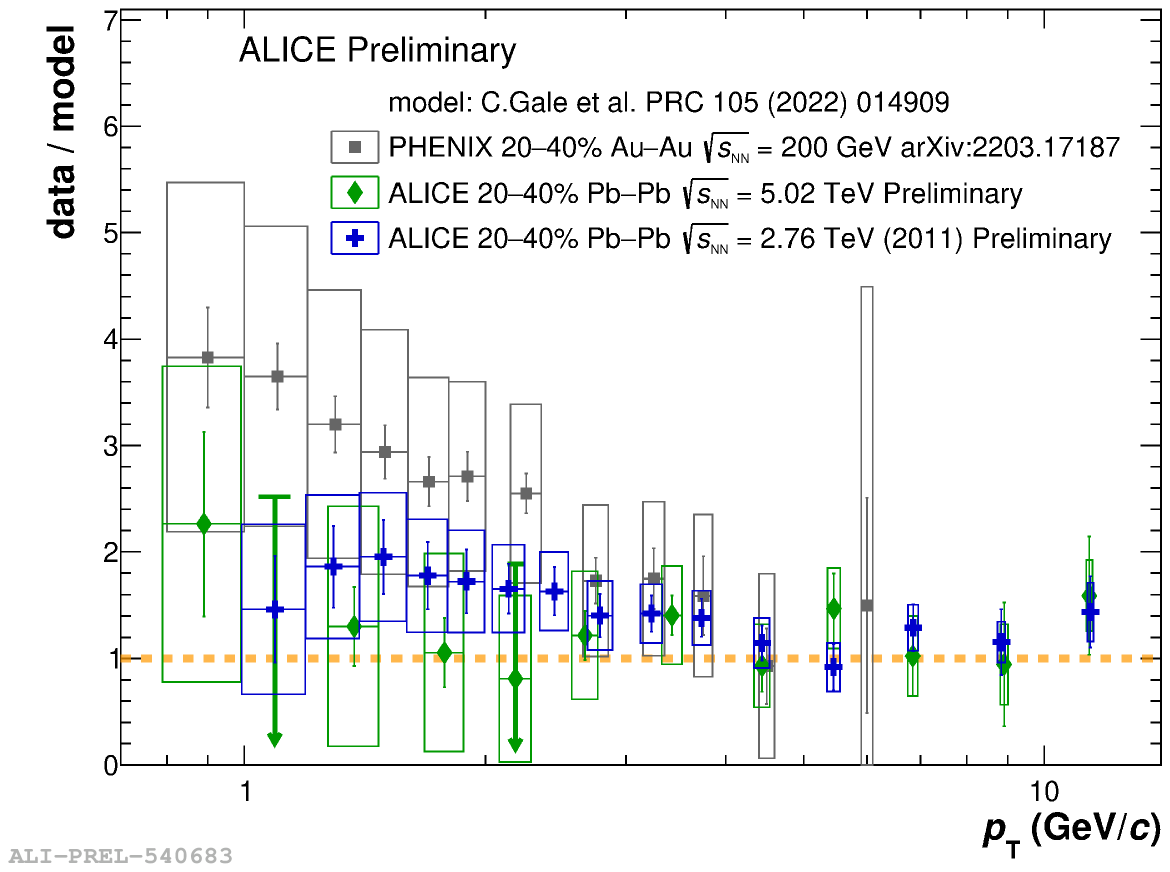}\\
\end{tabular}
\end{center}
 \caption{Ratios of direct-photon spectra to the expectations from the calculation of Ref.~\cite{Gale:2021emg}  for central (left) and semicentral (right) \AuAu and \PbPb collisions
at \sNN = 0.2, 2.76 and 5.02~TeV\xspace.}
  \label{fig:DirGammaModels2760}
\end{figure}
Direct-photon yields are measured subtracting from the inclusive-photon yields the estimated contribution from decay photons using a double ratio called \Rg, 
in order to reduce the systematic uncertainties; i.e.
$\gamma_{direct} = \gamma_{incl}- \gamma_{decay} = \gamma_{incl}  \cdot \left(1- \frac{1}{R_\gamma} \right)$, where $R_\gamma = \frac{(N_{\rm {\gamma, inc}}/N_{\pi^0})_{\rm measured} }{(N_{\rm {\gamma,decay}}/N_{\pi^0})_{\rm sim}}$.
The decay photon spectra are estimated employing the decay cocktail framework \cite{ALICE:2018mjj} using as input the measured \piz and \e meson spectra \cite{ALICE:2018mdl}  (or the \piz  spectra and \e/\piz ratio \cite{Danisch:2022m})  and \mT scaling for non measured spectra. 
The \Rg and corresponding  direct-photon spectra are shown in \hyperref[fig:Rgamma276]{Fig.~\ref*{fig:Rgamma276}}(left) and  \hyperref[fig:Rgamma276]{Fig.~\ref*{fig:Rgamma276}}(right) for central and semicentral \PbPb collisions at  \twosevensixnn.
In \hyperref[fig:Rgamma502]{Fig.~\ref*{fig:Rgamma502}}, the same observables are reported for  \fivenn.
In the \pT bins where the $\gamma_{\rm dir}$ is consistent with zero within total uncertainties, upper limited at 90\% CL are given.
At large \pT (\pT $>$ 4 \GeVc), the direct-photon signal is attributed to prompt photons, and agrees nicely with the scaled pQCD predictions for central and semicentral collisions at both energies.
In the intermediate \pT range (2-4 \GeVc), the direct-photon signals agree better with model predictions that include pre-equilibrium photons.
For  \pT $<$ 3 \GeVc  thermal-photon signals are observed in central and semicentral collisions at \twosevensixnn with  a significance of 3.1 (1.0 $<$ \pT $<$ 1.8 \GeVc) and 3.4 (1.0 $<$ \pT $<$ 2.3 \GeVc), respectively. 
At \fivenn no significant direct-photon signal  is observed at low \pT. 
Both results  are consistent with theoretical model predictions.
The direct-photon results at \twosevensixnn are in agreement with the previous publication \cite{Adam:2015lda} and show a larger significance 
due to the larger data sample and improved material budget uncertainties \cite{ALICE:2023kzv}. Furthermore, results in 0--10\% central \PbPb collisions could be obtained.
\begin{figure}[hbt]
\begin{center}
\begin{tabular}{cc}
\includegraphics[width=0.43\linewidth,trim={0.5cm 4.2cm 0.5cm 4.5cm}, clip ]{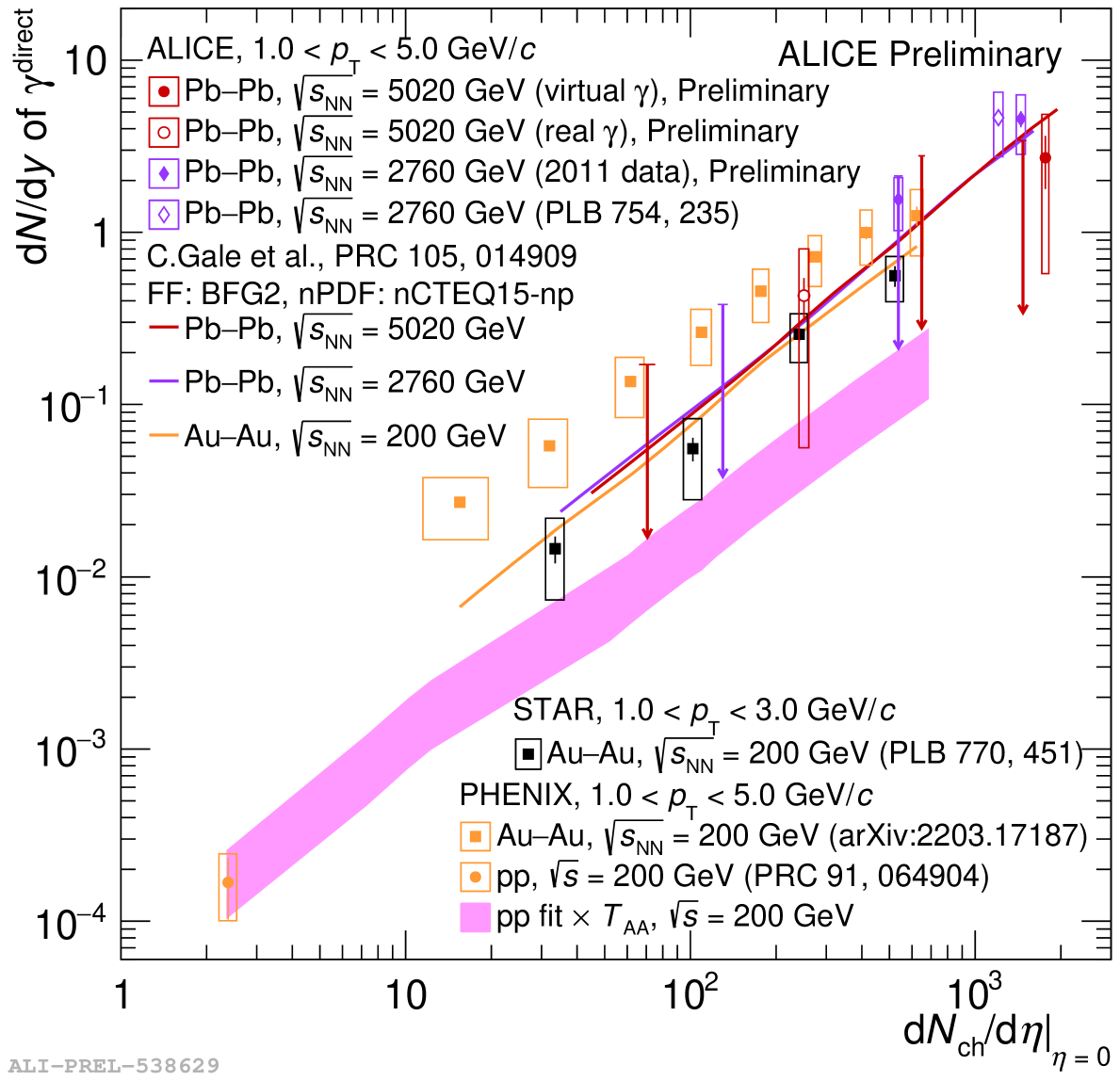}&
\includegraphics[width=0.42\linewidth,trim={0.1cm 1.9cm 0.1cm 1.9cm}, clip]{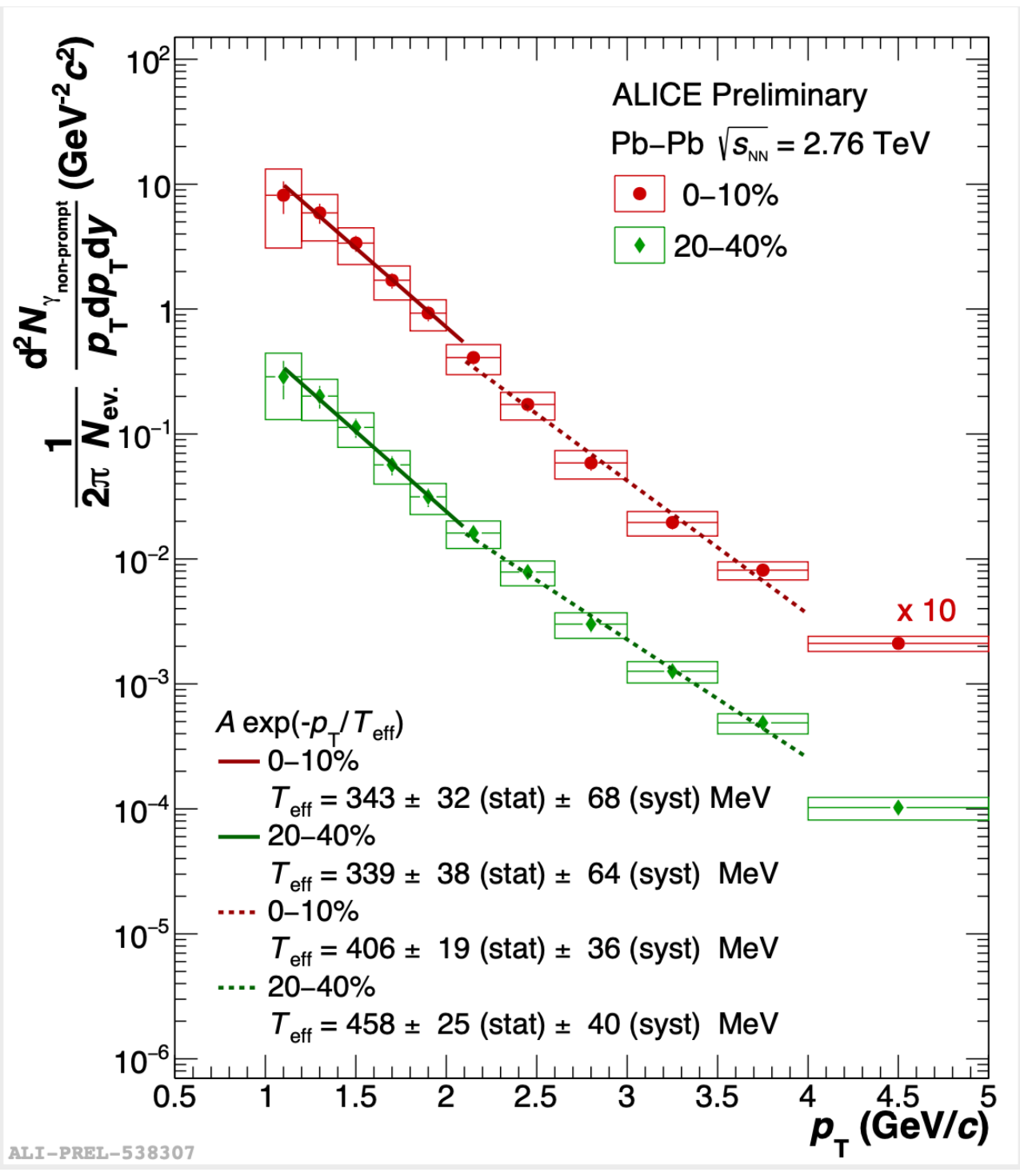}\\
\end{tabular}
\end{center}
 \caption{Left: Integrated direct $\gamma$ yields in the \pT range 1 \GeVc $<$ \pT $<$ 5\ GeVc as a function of the charged particle pseudorapidity density for different collision systems (pp, \AuAu and \PbPb)  
from \twoHnn to \fivenn  together with expectations from Ref.~\cite{Gale:2021emg}.
 Right: Non-prompt direct-photon spectra in central and semicentral \PbPb collisions at \twosevensixnn.}
  \label{fig:DirGammaTfit}
\end{figure}

Ratios of direct-photon spectra to the expectations from the theoretical model of Ref.~\cite{Gale:2021emg} are calculated  for central and semicentral \AuAu and \PbPb collisions
at \sNN = 0.2, 2.76 and 5.02 TeV \cite{PHENIX:2022rsx,Danisch:2022m} (see \hyperref[fig:DirGammaModels2760]{Fig.~\ref*{fig:DirGammaModels2760}}) with the aim of 
finding out whether state-of-the-art model calculations describe similarly direct-photon production at different energies.
A good agreement between ALICE data and model predictions is observed, while a slight tension may be present at low \pT  and semicentral collisions for the PHENIX data.
  The integrated direct $\gamma$ yields in the range 1 \GeVc $<$ \pT $<$ 5\ GeVc at both energies and different centralities 
  are plotted as a function of the charged particle pseudorapidity density together with the existing measurements \cite{PHENIX:2022rsx} in \hyperref[fig:DirGammaTfit]{Fig.~\ref*{fig:DirGammaTfit}}
    and  expectations from a model calculation \cite{Gale:2021emg}.
  A power-law scaling of direct $\gamma$ yields vs charged particle pseudorapidity density  is observed.
  The measured yields agree with the theoretical expectations at LHC energies. At RHIC energies, STAR results agree in all centralities with theoretical expectations while
  an agreement is observed only for  central and semicentral collisions for PHENIX results, with an increasing  difference between measured and predicted yields developing from semicentral to peripheral collisions.
    
The non-prompt direct-photon spectra at \twosevensixnn are obtained subtracting the scaled pQCD prompt photon contribution from the direct-photon spectrum in central and semicentral collisions at \twosevensixnn. The thermal spectra are 
then fitted with an exponential function in order to extract an effective temperature, $T_{\rm eff}$, of the medium  
(see \hyperref[fig:DirGammaTfit]{Fig.~\ref*{fig:DirGammaTfit}}).  
Exponential function fits in the low \pT range, 1.1 \GeVc $<$ \pT $<$ 2.1 \GeVc, deliver
$T_{\rm eff} = 343 \pm 32{\rm (stat)} \pm 68{\rm (sys)}$ MeV and   $T_{\rm eff}  = 339 \pm 32{\rm (stat)} \pm 68{\rm (sys)}$ MeV for central and semicentral collisions. 
These are the highest effective temperatures measured in heavy-ion collisions \cite{PHENIX:2022rsx,ALICE:2022wpn}. 
In  the  intermediate \pT range, 2.1 \GeVc $<$ \pT $<$ 4.0 \GeVc, the effective temperatures are $T_{\rm eff}  = 406 \pm 19{\rm (stat)} \pm 36{\rm (sys)}$ MeV and   $T_{\rm eff}  = 458 \pm 25{\rm (stat)} \pm 40{\rm (sys)}$ MeV. 
The  overall larger value of $T_{\rm eff}$  in the intermediate \pT range with respect to the values  in the low \pT range may be explained by an earlier emission time.



\section{Conclusions}
A first measurement of direct photons in \PbPb collisions at \fivenn is presented from central to peripheral collisions. Furthermore, new and improved measurements for central and semicentral collisions at \twosevensix
are also shown. Direct photon spectra at the LHC are in agreement with state-of the-art calculations. 

\bibliographystyle{JHEP}
\bibliography{ref}


\end{document}